\documentclass[conference]{IEEEtran}
\IEEEoverridecommandlockouts

\usepackage{cite}
\usepackage{amsmath,amssymb,amsfonts}
\usepackage[linesnumbered,ruled,vlined]{algorithm2e}
\usepackage{graphicx}
\usepackage{textcomp}
\usepackage{xcolor}
\usepackage{graphicx}
\usepackage{amsmath}
\usepackage[hidelinks]{hyperref}
\usepackage{orcidlink} 

\def\BibTeX{{\rm B\kern-.05em{\sc i\kern-.025em b}\kern-.08em
    T\kern-.1667em\lower.7ex\hbox{E}\kern-.125emX}}
\begin{document}

\title{KillChainGraph: ML Framework for Predicting and Mapping ATT\&CK Techniques\\}

\author{
\IEEEauthorblockN{Chitraksh Singh\,\orcidlink{0009-0000-1020-8989}} 
\IEEEauthorblockA{\textit{Frondeur Labs} \\
Mumbai, Maharashtra, INDIA \\
chitrakshsingh007@gmail.com}
\and
\IEEEauthorblockN{Monisha Dhanraj\, \orcidlink{0009-0009-8593-1632}}
\IEEEauthorblockA{\textit{Frondeur Labs} \\
Bengaluru, Karnataka, INDIA \\
monishadhanraj@frondeurlabs.com}
\and
\IEEEauthorblockN{Ken Huang\, \orcidlink{0009-0004-6502-3673}}
\IEEEauthorblockA{\textit{DistributedApps.ai, OWASP} \\
Fairfax, VA, USA\\
ken.huang@owasp.org}
}

\maketitle
\begin{abstract}
The escalating complexity and volume of cyberattacks demand proactive detection strategies that go beyond traditional rule-based systems. This paper presents a phase-aware, multi-model machine learning framework that emulates adversarial behavior across the seven phases of the Cyber Kill Chain using the MITRE ATT\&CK Enterprise dataset. Techniques are semantically mapped to phases via ATTACK-BERT, producing seven phase-specific datasets. We evaluate LightGBM, a custom Transformer encoder, fine-tuned BERT, and a Graph Neural Network (GNN), integrating their outputs through a weighted soft voting ensemble. Inter-phase dependencies are modeled using directed graphs to capture attacker movement from reconnaissance to objectives. The ensemble consistently achieved the highest scores, with F1-scores ranging from 97.47\% to 99.83\%, surpassing GNN performance (97.36\% to 99.81\%) by 0.03\%–0.20\% across phases. This graph-driven, ensemble-based approach enables interpretable attack path forecasting and strengthens proactive cyber defense.
\end{abstract}


\section{Introduction}

Cyberattacks have emerged as a significant threat to national security, critical infrastructure, and global economic stability. According to the United Nations Office on Drugs and Crime (UNODC), cybercrime inflicts over \$6 trillion in annual damages globally, with projections indicating a sharp increase by 2025~\cite{unodc2020cyber}. The IBM Cost of a Data Breach Report 2023 places the average cost of a breach at \$4.45 million, underscoring the rising sophistication and impact of modern cyber threats targeting high-value domains such as finance, healthcare, and government systems~\cite{ibm2023cost}.

A foundational lens for understanding adversarial behavior is the Cyber Kill Chain framework introduced by Lockheed Martin~\cite{hutchins2011intelligence}. This model dissects cyber intrusions into seven sequential stages: Reconnaissance, Weaponization, Delivery, Exploitation, Installation, Command and Control (C2), and Actions on Objectives. Alternative frameworks such as the MITRE ATT\&CK framework~\cite{strom2018mitre}, the Diamond Model of Intrusion Analysis~\cite{caltagirone2013diamond}, and the Unified Kill Chain~\cite{hubbard2017unified} have been developed to address limitations in linear or high-level representations. These models incorporate more granular attacker tactics, behavioral patterns, and infrastructure elements. Other notable models include the Mandiant Attack Lifecycle~\cite{mandiant2013attack}, which focuses on response and remediation, the OODA Loop (Observe–Orient–Decide–Act) applied in cyber operations for dynamic decision making, and the MITRE Engage framework~\cite{mitreengage2021}, which supports active defense and deception strategies. Additionally, the Extended Cyber Kill Chain (ECKC)~\cite{hutchins2014eckc} expands upon the traditional model by incorporating internal reconnaissance, lateral movement, and privilege escalation more explicitly.

Conventional security tools such as firewalls, rule-based intrusion detection systems, and signature based analysis \cite{ddos} operate reactively and are often ineffective against zero day vulnerabilities or polymorphic malware. Machine Learning (ML), on the other hand, offers scalable solutions for extracting behavioral insights from large volumes of cyber threat intelligence data~\cite{sommer2010outside, shaukat2020survey}. Recent works have shown success in leveraging ML for intrusion detection and anomaly classification, but very few have focused on phase-specific modeling aligned with structured frameworks like MITRE ATT\&CK and CKC.

In this work, we present a phase-aware machine learning framework for modeling adversarial techniques across the cyber kill chain using the MITRE ATT\&CK knowledge base. Seven curated phase-specific datasets are constructed by semantically aligning technique descriptions with their respective phases. Four classifiers LightGBM, a custom Transformer encoder, fine-tuned BERT, and a Graph Neural Network (GNN) are trained to predict likely adversarial techniques within each phase. Their outputs are combined using a weighted soft voting ensemble to improve predictive accuracy. Inter-phase dependencies are modeled through directed graphs linking earlier-phase predictions (e.g., T1595, T1087 in Reconnaissance) to subsequent phases, producing interpretable attack paths that enhance defender situational awareness.

The GNN classifier yielded superior performance, achieving 99.28\% accuracy in the Delivery phase and above 97.35\% across all others. Fine-tuned BERT exhibited strong results, while LightGBM achieved F1-scores between 0.91 and 0.96. The custom Transformer reached up to 87.00\% accuracy in \textit{Actions on Objectives} and 86.81\% in F1-score, though it underperformed in low-sample contexts such as \textit{Delivery} with 55.56\% accuracy. These results validate the effectiveness of our phase-wise, graph-driven architecture.

\begin{table*}[ht]
    \centering
    \caption{Table of Cyber Kill Chain Modeling Approaches}
    \resizebox{\textwidth}{!}{%
    \large
    \begin{tabular}{|l|l|l|l|c|c|c|}
        \hline
        \textbf{Author (Year)} & \textbf{Approach} & \textbf{Key Contributions} & \textbf{Limitations} & \textbf{Phase-wise Modeling} & \textbf{Semantic Linking} & \textbf{ML-Based Prediction} \\
        \hline
        Hutchins et al. (2011) \cite{hutchins2011intelligence} & Cyber Kill Chain & Defined 7-stage attacker lifecycle & No automation or technique-level modeling & \checkmark & X & X \\
        \hline
        Strom et al. (2018) \cite{strom2018mitre} & MITRE ATT\&CK & Curated real-world adversarial techniques & Lacks sequential kill chain structure & X & X & X \\
        \hline
        Nguyen et al. (2018) \cite{mlkillchain} & Cyber Kill Chain for Attacking ML Models & Novel kill chain tailored to ML attacks & Not designed for general adversarial detection or defense modeling & \checkmark & X & X \\
        \hline
        Boltan et al. (2023) \cite{attackknowgraph} & ATT\&CK Knowledge Graph (KG) & Investigated ATT\&CK representation in knowledge graphs & Focused on representation, lacks predictive modeling of attack flow & X & \checkmark & X \\
        \hline
        Meng et al. (2023) \cite{gnnkillchain} & GNN for Attack Path Detection & Novel GNN algorithm for detecting paths across attack graphs & Not explicitly aligned with known kill chain frameworks & \checkmark & X & \checkmark \\
        \hline
    \end{tabular}%
    }
    \label{tab:kill_chain_modeling}
\end{table*}

\section{Related Work}

Modeling adversarial behavior through structured frameworks has long been central to cybersecurity research. The Cyber Kill Chain model, introduced by Hutchins et al.~\cite{hutchins2011intelligence}, presents a seven-stage attack lifecycle that captures the sequential flow of an intrusion, from reconnaissance to final exploitation. It became foundational for both blue and red team operations. Later, the MITRE ATT\&CK framework~\cite{strom2018mitre} offered a granular view of real-world techniques and tactics observed in advanced persistent threats, enabling greater fidelity in threat detection and red teaming exercises. 

Nguyen et al.~\cite{mlkillchain} introduced a kill chain specific to attacking machine learning models, but focused more on theoretical modeling than actionable defense. Boltan et al.~\cite{attackknowgraph} explored knowledge graph representations for MITRE ATT\&CK but did not address predictive capabilities across phases. Meng et al.~\cite{gnnkillchain} used graph neural networks to identify attack paths, though their work lacked alignment with structured kill chain phases. Recent studies have also utilized dynamic graph construction to trace multi-stage intrusions, capturing evolving relationships among tactics and techniques in real time.

In contrast, our approach introduces a forward-predictive pipeline that models attacker progression across the cyber kill chain using a suite of supervised classifiers LightGBM, Transformer, BERT, and GNN each trained per kill chain phase. The input dataset was curated by semantically aligning MITRE ATT\&CK techniques with the Lockheed Martin kill chain phases using ATTACK-BERT. This generated phase-specific datasets that enable fine-grained classification of adversarial behavior. A semantic similarity graph is then constructed to link predicted techniques across phases, effectively simulating the chaining logic employed by threat actors. This allows defenders to proactively identify potential full-cycle attack paths and preemptively strengthen control mechanisms.
Unlike prior works that either visualize attack flows or statically map adversary behavior, our method provides a dynamic, data-driven framework for predicting chained attack stages. Furthermore, the combination of semantically guided data engineering and graph-based inference enables explainable reasoning over adversarial tactics something largely missing in existing detection-focused research. The integration of transformer-based models with contextual graphs represents a novel direction in cyber threat modeling.

Table~\ref{tab:kill_chain_modeling} presents a comparative summary of key literature in this domain, highlighting the novelty of our framework in combining semantically guided ML classification with graph-based reasoning over kill chain stages.

\section{Methodology}

\begin{figure}
    \centering
    \includegraphics[width=1\linewidth]{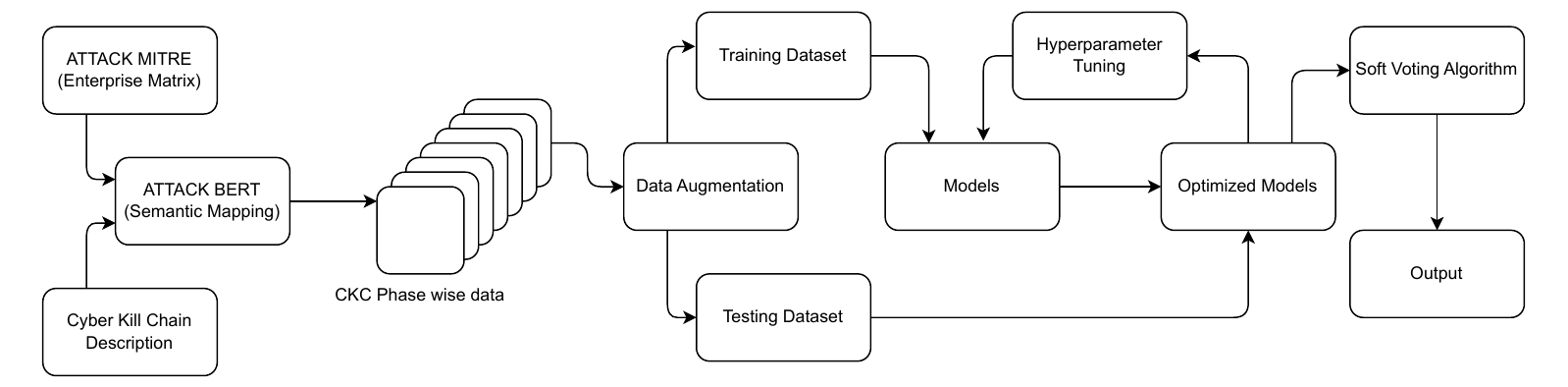}
    \caption{Model Training Process Flow for predicting ATT\&CK technique}
    \label{fig:process_flow}
\end{figure}

This section outlines the methodology adopted for generating cyber kill chains using machine learning. The process consists of two major components: dataset construction and model training as shown in Fig. \ref{fig:process_flow}.

\subsection{Dataset}

The dataset used in this study was constructed by semantically aligning adversarial behavior descriptions from the MITRE ATT\&CK Enterprise matrix with the Lockheed Martin Cyber Kill Chain (CKC) framework. Using ATTACK-BERT for semantic similarity, each ATT\&CK technique was mapped to an appropriate CKC phase, forming a comprehensive dataset of labeled adversarial actions. Each sample in the dataset consists of a natural language description of a technique (\texttt{combine\_description}), its corresponding ATT\&CK technique ID and name (\texttt{Technique\_Name}), and the associated kill chain phase. This semantic mapping yielded a unified dataset that was subsequently split into seven distinct phase-specific datasets, corresponding to Reconnaissance, Weaponization, Delivery, Exploitation, Installation, Command \& Control, and Objectives.

The raw descriptions were preprocessed by removing special characters, normalizing case, and cleaning noise, ensuring that input text was standardized and ready for downstream modeling. Each of the seven datasets was stratified into training ,validation and testing sets using an 70-10-20 split, with an emphasis on maintaining class balance wherever feasible.

To address class imbalance and limited data availability especially in phases like Delivery and Command \& Control a diverse set of text data augmentation strategies was applied. Synonym substitution was conducted using both lexical resources like WordNet and contextual language models such as BERT via the \texttt{nlpaug} library. Additionally, TF-IDF based token filtering was employed to drop low-importance terms, helping models focus on domain-relevant features. Sentence-level variability was introduced through operations like word reordering and controlled duplication, which enhanced linguistic diversity without altering the underlying intent. To further enrich the training corpus, paraphrasing techniques were applied using Pegasus-based generative models and multilingual back-translation via MarianMT, utilizing intermediate languages such as French, Spanish, and German.

Together, these augmentation strategies significantly improved the robustness and generalization ability of models trained on low-resource kill chain phases. The final phase-wise datasets thus provided a solid foundation for training specialized classifiers, enabling precise prediction of ATT\&CK techniques based on adversarial behavior descriptions across different stages of the cyber kill chain.

\subsection{LGBM Model Architecture}

The LightGBM (LGBM) model was employed as an efficient and scalable gradient boosting framework for classifying ATT\&CK techniques. The objective was to learn a phase-specific mapping from sentence embeddings to the corresponding ATT\&CK technique labels. Given an input dataset \( \mathcal{D} = \{(x_1, y_1), \dots, (x_n, y_n)\} \), where each \( x_i \) represents a sentence embedding generated from a \texttt{combine\_description} field and \( y_i \in \mathcal{T} \) denotes its corresponding technique label, the LGBM classifier learns a function \( f: \mathbb{R}^d \rightarrow \mathcal{T} \) as given in Algorithm~\ref{alg:lgbm_classifier}.

\begin{algorithm}
\DontPrintSemicolon
\KwIn{Set of descriptions $\{x_i\}_{i=1}^n$, labels $\{y_i\}_{i=1}^n$, pre-trained encoder $\mathcal{E}$}
\KwOut{Trained LGBM model $f^*(x)$}

\textbf{LGBM Training Process:} \

Initialize empty embedding matrix $X$ and label vector $Y$ \;

\ForEach{sample $x_i$}{
    Compute embedding: $\mathbf{e}_i \leftarrow \mathcal{E}(x_i)$ \;
    
    Append $\mathbf{e}_i$ to $X$ and $y_i$ to $Y$ \;
}

Split $(X, Y)$ into training and testing sets: $(X_{\text{train}}, Y_{\text{train}}, X_{\text{test}}, Y_{\text{test}})$ \;

Configure LGBMClassifier with hyperparameters (e.g., \texttt{num\_leaves}, \texttt{learning\_rate}, \texttt{max\_depth}, \texttt{early\_stopping\_rounds}) \;

Train LGBM on $(X_{\text{train}}, Y_{\text{train}})$ with early stopping using $(X_{\text{test}}, Y_{\text{test}})$ \;

Evaluate model on test set using multi-class log loss objective fucntion Eq.\ref{eq:loss_function}

Return trained model $f^*(x)$ \;

\caption{LGBM-based ATT\&CK Technique Classification}
\label{alg:lgbm_classifier}
\end{algorithm}

Embeddings were generated using pre-trained models such as ATTACK-BERT, resulting in fixed-length vectors \( \mathbf{E}(x_i) \in \mathbb{R}^d \), which served as input features. The LGBM model then trained decision trees in a leaf wise manner using histogram based gradient boosting. Hyperparameters like the number of leaves, maximum depth, learning rate, and early stopping rounds were tuned using grid search and cross-validation. Class imbalance was addressed using class weight balancing.

The objective function minimized was the multi-class log loss:

\begin{equation}
\mathcal{L} = -\frac{1}{n} \sum_{i=1}^{n} \log P(y_i \mid \mathbf{E}(x_i); \theta),
\label{eq:loss_function}
\end{equation}

where \( P(y_i \mid \mathbf{E}(x_i); \theta) \) denotes the class probability predicted by the model with parameters \( \theta \).

\subsection{Transformer-based Classifier}

The Transformer-based classifier leverages the self-attention mechanism to model long-range dependencies and contextual information in the input textual descriptions. Each ATT\&CK technique description is tokenized and embedded into a dense vector space using pre-trained GloVe embeddings. These embeddings serve as the input to a Transformer encoder model designed to classify the technique name associated with a given description.

Formally, let $X = \{x_1, x_2, \dots, x_n\}$ represent a tokenized description of length $n$. Each token $x_i$ is mapped to a $d$-dimensional vector via an embedding matrix $E \in \mathbb{R}^{|V| \times d}$, where $V$ is the vocabulary. Positional encoding is added to retain the sequential order, producing the input matrix $X' \in \mathbb{R}^{n \times d}$. This is passed through a stack of Transformer encoder blocks, each comprising Multi-Head Self-Attention (MHSA), layer normalization, and position-wise feed-forward networks (FFN).

The output of the final encoder layer is passed through a global average pooling layer to obtain a fixed-size representation, followed by a linear classification head with a softmax activation. The model is trained using cross-entropy loss as given in Algorithm~\ref{alg:transformer_classifier}.

\begin{algorithm}
\DontPrintSemicolon
\KwIn{Technique description $T$, GloVe embeddings $E_{\text{glove}}$}
\KwOut{Predicted label $\hat{y}$}

\textbf{Transformer-based Text Classification Process:} \

Tokenize the input $T$ into sequence of tokens $\{x_1, x_2, \dots, x_n\}$ \;

Embed tokens using GloVe: $X \leftarrow E_{\text{glove}}(x_1, \dots, x_n)$ \;

Add positional encoding: $X' \leftarrow X + PE$ \;

\ForEach{transformer encoder layer $l$}{
    a. Apply multi-head self-attention: $Z_l \leftarrow \text{MHSA}(X')$ \;

    b. Add residual connection and layer normalization: $X' \leftarrow \text{LayerNorm}(Z_l + X')$ \;

    c. Apply feed-forward network, residual connection, and normalization: $X' \leftarrow \text{LayerNorm}(\text{FFN}(X') + X')$ \;
}

Aggregate contextual embeddings: $h \leftarrow \text{GlobalAvgPool}(X')$ \;

Compute class probabilities: $\hat{y} \leftarrow \text{Softmax}(W h + b)$ \;

Return predicted label $\hat{y}$ \;

\caption{Transformer-based ATT\&CK Technique Classification}
\label{alg:transformer_classifier}
\end{algorithm}

This approach benefits from the inductive bias of self-attention and the flexibility of non-recurrent computation. Despite having fewer trainable parameters than large language models, the Transformer encoder demonstrated robust generalization on structured ATT\&CK texts.

\subsection{BERT-based Classifier}

BERT (Bidirectional Encoder Representations from Transformers) is a pre-trained language model designed to capture deep bidirectional representations from unlabeled text by jointly conditioning on both left and right context. In this classification task, we fine-tune a pre-trained \texttt{bert base uncased} model on ATT\&CK technique descriptions to predict the corresponding technique label.

Each input description is first tokenized [CLS] using the WordPiece tokenizer and truncated [SEP] or padded to a fixed sequence length. The tokens are passed through the BERT encoder, which outputs contextual embeddings for each token. The token’s embedding, which is designed to represent the entire sequence, is extracted and passed through a fully connected linear layer followed by a softmax function for classification as given in Algorithm~\ref{alg:bert_classifier}.

\begin{algorithm}
\DontPrintSemicolon
\KwIn{Technique description $T$, pre-trained BERT model $\mathcal{B}$}
\KwOut{Predicted label $\hat{y}$}

\textbf{BERT-based Text Classification Process:} \

Tokenize input $T$ using BERT tokenizer with special tokens \texttt{[CLS]} and \texttt{[SEP]} \;

Convert tokens to input IDs and generate attention mask \;

Feed inputs to BERT: $H \leftarrow \mathcal{B}(T)$ \;

Extract \texttt{[CLS]} token embedding: $h_{\text{cls}} \leftarrow H[0]$ \;

Compute class probabilities: $\hat{y} \leftarrow \text{Softmax}(W h_{\text{cls}} + b)$ \;

Return predicted label $\hat{y}$ \;

\caption{BERT-based ATT\&CK Technique Classification}
\label{alg:bert_classifier}
\end{algorithm}

BERT’s strength lies in its ability to learn task-specific representations through minimal architectural modification and fine-tuning. This allows the model to adapt to domain-specific text, such as structured adversarial descriptions in cybersecurity. Compared to standard Transformer encoders trained from scratch, BERT provides a performance boost due to its pre-training on large-scale corpora like Wikipedia and BookCorpus.

\subsection{GNN-based Classifier}

Graph Neural Networks (GNNs) are designed to operate on graph-structured data, capturing both node features and topological relationships. In our methodology, a GNN-based classifier is used to model the semantic and structural connections between ATT\&CK technique descriptions. Each description is treated as a node in the graph, and edges are constructed based on textual similarity (e.g., cosine similarity of embeddings or shared keyword patterns).

The node features are initialized using BERT-based sentence embeddings. The graph is constructed such that edges reflect high semantic similarity between techniques. A GNN layer, such as GraphSAGE or GAT, aggregates information from neighboring nodes to update the representation of each node. The final node embedding is passed through a classification layer to predict the corresponding technique label as given in Algorithm~\ref{alg:gnn_classifier}.

\begin{algorithm}
\DontPrintSemicolon
\KwIn{Technique nodes $\{v_i\}$ with embeddings $\{x_i\}$, graph $G = (V, E)$}
\KwOut{Predicted labels $\{\hat{y}_i\}$ for each node $v_i$}

\textbf{GNN-based Text Classification Process:} \

Construct graph $G$ based on semantic similarity between technique descriptions \;

Initialize node features $x_i$ using BERT embeddings \;

\ForEach{GNN layer $l = 1$ to $L$}{
    \ForEach{node $v_i \in V$}{
        3a. Aggregate neighbor features: \\
        \quad $m_i^{(l)} \leftarrow \text{AGGREGATE}^{(l)}(\{x_j^{(l-1)} \mid j \in \mathcal{N}(i)\})$ \;

        3b. Update node representation: \\
        \quad $x_i^{(l)} \leftarrow \text{COMBINE}^{(l)}(x_i^{(l-1)}, m_i^{(l)})$ \;
    }
}

Apply MLP classifier to final node embeddings: \\
\quad $\hat{y}_i \leftarrow \text{Softmax}(W x_i^{(L)} + b)$ \;

Return predicted labels $\{\hat{y}_i\}$ \;

\caption{GNN-based ATT\&CK Technique Classification}
\label{alg:gnn_classifier}
\end{algorithm}

The use of GNN allows the model to exploit the relational structure among technique descriptions, improving generalization for similar but rare classes. This is especially valuable in cybersecurity, where techniques may have overlapping semantics and evolving terminology.

\subsection{Ensemble Strategy}
To synthesize the predictions from our four trained models (LightGBM, custom Transformer, BERT, and GNN), we implemented a \textit{Weighted Soft Voting} ensemble strategy. Unlike hard voting, which relies on a simple majority vote of predicted labels, soft voting considers the predicted class probabilities from each classifier, enabling more nuanced decisions influenced by each model's confidence. The final prediction is determined by selecting the class label $c$ with the highest weighted average probability across all $M$ classifiers. The probability for each class $c$ is computed as follows:

\begin{equation}
P(c|x) = \sum_{i=1}^{M} \left( w_i \cdot p_i(c|x) \right)
\end{equation}

where $P(c|x)$ is the final ensemble probability for class $c$ given input $x$, $M$ is the number of classifiers in the ensemble (in our case, $M=4$), $w_i$ is the weight assigned to the $i$-th classifier, and $p_i(c|x)$ is the probability predicted by the $i$-th classifier for class $c$. The weights $w_i$ are critical for optimizing the ensemble's performance. Instead of assigning them equally, we determined the weights based on each model's performance on the validation set for each specific Cyber Kill Chain (CKC) phase. Specifically, the weight for each classifier was made proportional to its macro-averaged F1-score for that phase. This ensures that models demonstrating superior performance have a greater influence in the final decision, thereby enhancing predictive accuracy.

\subsection{Semantic Mapping and Graph Construction}

Once predictions from each phase-wise model are obtained consisting of technique labels (e.g., “Phishing”, “Spearphishing Link”) and their descriptions the next step is to construct a semantic graph that connects techniques across adjacent phases of the cyber kill chain. The goal is to simulate how an attacker might logically transition from one phase to the next.

Each predicted technique description is first converted into a dense vector representation using ATTACK-BERT, a BERT-based model fine-tuned on cybersecurity specific language. These vectors capture the semantic meaning of each technique. For any two techniques $v_i$ (from phase $t$) and $v_j$ (from phase $t+1$), we compute how similar they are using cosine similarity, defined as:

\begin{equation}
\text{sim}(v_i, v_j) = \frac{v_i \cdot v_j}{\|v_i\| \|v_j\|}
\end{equation}

Here, $v_i \cdot v_j$ denotes the dot product of the two vectors, and $\|v_i\|$ and $\|v_j\|$ are their magnitudes. Cosine similarity ranges from -1 (completely opposite) to 1 (perfectly similar). In our case, values closer to 1 indicate that the techniques are semantically similar and might represent a logical transition in an attacker’s plan.

To build the graph, we use a threshold $\tau$ (typically between 0.7 and 0.9). If the similarity between two techniques across adjacent phases is greater than or equal to $\tau$, we draw a directed edge from the earlier-phase technique to the later-phase technique. This creates a semantic graph $G = (V, E)$, where:
$V$ is the set of all predicted techniques from all phases,
$E$ is the set of edges linking techniques across phases based on semantic similarity. This process is outlined in Algorithm~\ref{alg:semantic_mapping}.

\begin{algorithm}
\DontPrintSemicolon
\KwIn{Predicted technique labels $\{v_i^t\}$ for each phase $P_t$, corresponding description embeddings $\{x_i^t\}$ from ATTACK-BERT, similarity threshold $\tau$}
\KwOut{Semantic graph $G = (V, E)$ connecting cross-phase technique nodes}

Initialize empty directed graph: $G = (V, E)$ \;

\ForEach{phase $t = 1$ to $n - 1$}{
    \ForEach{node $v_i^t \in P_t$}{
        \ForEach{node $v_j^{t+1} \in P_{t+1}$}{
            2a. Compute cosine similarity between embeddings: \\
            \quad $s_{ij} = \frac{x_i^t \cdot x_j^{t+1}}{\|x_i^t\|\|x_j^{t+1}\|}$ \;

            2b. \If{$s_{ij} \geq \tau$}{
                Add directed edge $v_i^t \rightarrow v_j^{t+1}$ to $E$ \;
            }
        }
    }
}

Return semantic graph $G = (V, E)$ \;

\caption{Semantic Graph Construction}
\label{alg:semantic_mapping}
\end{algorithm}

This semantic mapping helps model an attacker’s potential movement across phases e.g., a reconnaissance technique being logically followed by a specific weaponization technique. The resulting graph provides an interpretable and structured view of predicted attacker behavior, aiding in downstream visualization, reasoning, or alert prioritization.

\section{Experimental Setup}

All experiments were conducted on an NVIDIA Tesla P100 GPU with 16GB memory, offering sufficient computational power for deep learning model training across multiple datasets, each corresponding to a distinct phase of the cyber kill chain (CKC). By segmenting the MITRE ATT\&CK data according to these phases, each model was trained independently to specialize in detecting adversarial techniques relevant to its respective CKC stage. This phase-wise strategy allowed the system to simulate the sequential nature of attacker behaviors.

The primary task was to predict the appropriate MITRE ATT\&CK techniques for each CKC phase based on threat report descriptions. The output of each phase's classifier was semantically linked to subsequent phases using cosine similarity of embedding vectors, forming a directed graph to model the progression of adversarial actions.

A variety of models were employed to cover different algorithmic paradigms. These included BERT-base-uncased model fine-tuned with cross-entropy loss, Transformer-based classifiers with attention mechanisms and positional encodings, ensemble LightGBM classifiers trained on sentence embeddings, and GNN-based models leveraging co-occurrence windows and word embeddings for graph construction.

\begin{table*}[htbp]
\centering
\caption{Performance Comparison of LGBM, GNN, Transformer, BERT, and Ensemble Models on test dataset}
\label{tab:ensemble_results}
\begin{tabular}{|l|l|c|c|c|c|}
\hline
\textbf{Phase} & \textbf{Model} & \textbf{Accuracy (\%)} & \textbf{Precision (\%)} & \textbf{Recall (\%)} & \textbf{F1-Score (\%)} \\
\hline
Reconnaissance
& LGBM         & 93.00 & 93.00 & 92.00 & 91.00 \\
\cline{2-6}
& GNN          & 97.35 & 97.40 & 97.33 & 97.36 \\
\cline{2-6}
& Transformer  & 77.78 & 78.00 & 77.50 & 77.74 \\
\cline{2-6}
& BERT         & 80.67 & 81.00 & 80.00 & 80.49 \\
\cline{2-6}
& Ensemble     & \textbf{98.35} & \textbf{97.43} & \textbf{97.36} & \textbf{97.50}\\
\hline

Weaponization
& LGBM         & 96.00 & 97.00 & 96.00 & 96.00 \\
\cline{2-6}
& GNN          & 98.83 & 98.85 & 98.81 & 98.83 \\
\cline{2-6}
& Transformer  & 75.28 & 75.50 & 75.00 & 75.25 \\
\cline{2-6}
& BERT         & 79.41 & 79.50 & 79.00 & 79.24 \\
\cline{2-6}
& Ensemble     & \textbf{98.86} & \textbf{98.88} & \textbf{99.81} & \textbf{98.86} \\
\hline

Delivery
& LGBM         & 88.00 & 88.00 & 87.00 & 88.00 \\
\cline{2-6}
& GNN          & 99.28 & 99.30 & 99.27 & 99.28 \\
\cline{2-6}
& Transformer  & 55.56 & 56.00 & 55.00 & 55.49 \\
\cline{2-6}
& BERT         & 89.50 & 90.00 & 89.00 & 89.49 \\
\cline{2-6}
& Ensemble     & \textbf{99.31} & \textbf{99.34} & \textbf{99.30} & \textbf{99.31} \\
\hline

Exploitation
& LGBM         & 93.00 & 93.00 & 92.00 & 91.00 \\
\cline{2-6}
& GNN          & 97.66 & 97.71 & 97.64 & 97.67 \\
\cline{2-6}
& Transformer  & 79.84 & 80.00 & 79.50 & 79.74 \\
\cline{2-6}
& BERT         & 82.04 & 82.50 & 82.00 & 82.24 \\
\cline{2-6}
& Ensemble     & \textbf{98.66} & \textbf{97.74} & \textbf{97.66} & \textbf{97.87}\\
\hline

Installation
& LGBM         & 95.00 & 94.00 & 95.00 & 94.00 \\
\cline{2-6}
& GNN          & 98.69 & 98.72 & 98.68 & 98.70 \\
\cline{2-6}
& Transformer  & 76.56 & 77.00 & 76.00 & 76.49 \\
\cline{2-6}
& BERT         & 79.75 & 80.00 & 79.50 & 79.74 \\
\cline{2-6}
& Ensemble     & \textbf{98.72} & \textbf{99.72} & \textbf{98.71} & \textbf{98.83}\\
\hline

Command \& Control
& LGBM         & 94.00 & 94.00 & 94.00 & 93.00 \\
\cline{2-6}
& GNN          & 97.43 & 97.48 & 97.41 & 97.44 \\
\cline{2-6}
& Transformer  & 64.62 & 65.00 & 64.00 & 64.49 \\
\cline{2-6}
& BERT         & 78.73 & 79.00 & 78.50 & 78.74 \\
\cline{2-6}
& Ensemble     & \textbf{98.43} & \textbf{97.50} & \textbf{97.44} & \textbf{97.47} \\
\hline

Actions on Objectives
& LGBM         & 93.00 & 93.00 & 93.00 & 93.00 \\
\cline{2-6}
& GNN          & 98.77 & 98.80 & 98.75 & 98.77 \\
\cline{2-6}
& Transformer  & 86.81 & 87.00 & 86.50 & 86.74 \\
\cline{2-6}
& BERT         & 84.33 & 84.50 & 84.00 & 84.24 \\
\cline{2-6}
& Ensemble     & \textbf{99.77} & \textbf{98.83} & \textbf{98.78} & \textbf{98.80} \\
\hline
\end{tabular}
\end{table*}

\subsection{Model Training}

Each CKC phase dataset was used to train a dedicated classifier to predict MITRE ATT\&CK techniques from composite descriptions. We explored four primary approaches: a BERT-based classifier, a Transformer-based neural classifier, a LightGBM ensemble model, and a graph-based GNN classifier.

For the BERT-based model, we fine-tuned the \texttt{bert base uncased} variant using the HuggingFace Transformers library. The input texts were tokenized with a maximum sequence length of 128 and padded accordingly. Training was performed for 50 epochs using a batch size of 32 and a learning rate of $2\mathrm{e}^{-5}$, optimizing cross-entropy loss via the AdamW optimizer. The model's architecture was left unchanged except for the final classification head, which was adapted to the number of class labels for each phase. Stratified train-test splits (80/20) were used for evaluation.

The Transformer-based classifier was implemented using PyTorch with custom positional encoding and transformer encoder layers. Tokens were embedded using a learnable embedding layer of dimension 128, passed through a positional encoding block, and processed via a 2-layer Transformer encoder with 4 attention heads and feedforward size of 256. Mean pooling was applied across the sequence to derive fixed-length representations, followed by a linear classification layer. Each phase model was trained independently for 50 epochs with a batch size of 32 and learning rate of $1\mathrm{e}^{-4}$ using the Adam optimizer. The vocabulary size matched the tokenizer from \texttt{bert base uncased}, and gradients were clipped to 0.5 to stabilize training. Training and validation losses were tracked, and loss plots were saved to monitor convergence.

The LightGBM model was trained on concatenated sentence-transformer embeddings. Hyperparameter tuning included number of leaves ranging from 31 to 100, learning rates of \{0.01, 0.05, 0.1, 0.2\}, estimators ranging from 100 to 400, L2 regularization $1\mathrm{e}^{-8}$, and maximum depths from 5 to 25.

For the GNN classifier, documents were converted into graph structures where nodes represented unique words and edges denoted co-occurrence within a sliding window. Word embeddings were initialized using GloVe vectors of 50 dimensions, with out-of-vocabulary words assigned random vectors. Graph construction leveraged DGL, and node-level features were processed using two Graph Convolution layers followed by average pooling. The model was optimized with AdamW, using a learning rate of $1\mathrm{e}^{-3}$, hidden dimension of 64, and batch size of 32. Evaluation was performed using stratified 3-fold cross-validation.

\begin{figure*}
    \centering
    \includegraphics[width=1\linewidth]{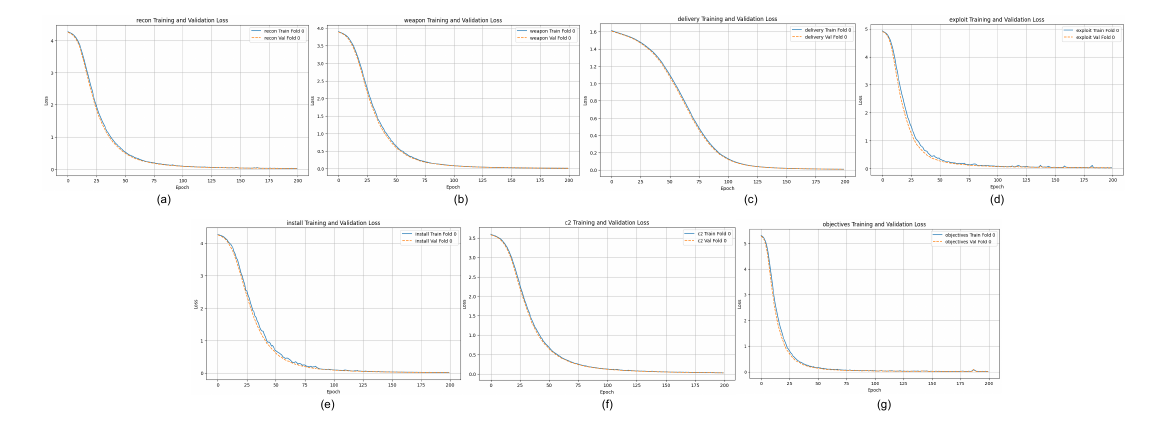}
    \caption{Training and validation loss graph for each GNN model based on the CKC phase: (a) Reconnaissance, (b) Weaponization,
(c) Delivery, (d) Exploitation, (e) Installation, (f) Command \& Control, and (g) Objectives.}
    \label{fig:gnn_loss}
\end{figure*}

\begin{figure*}
    \centering
    \includegraphics[width=1\linewidth]{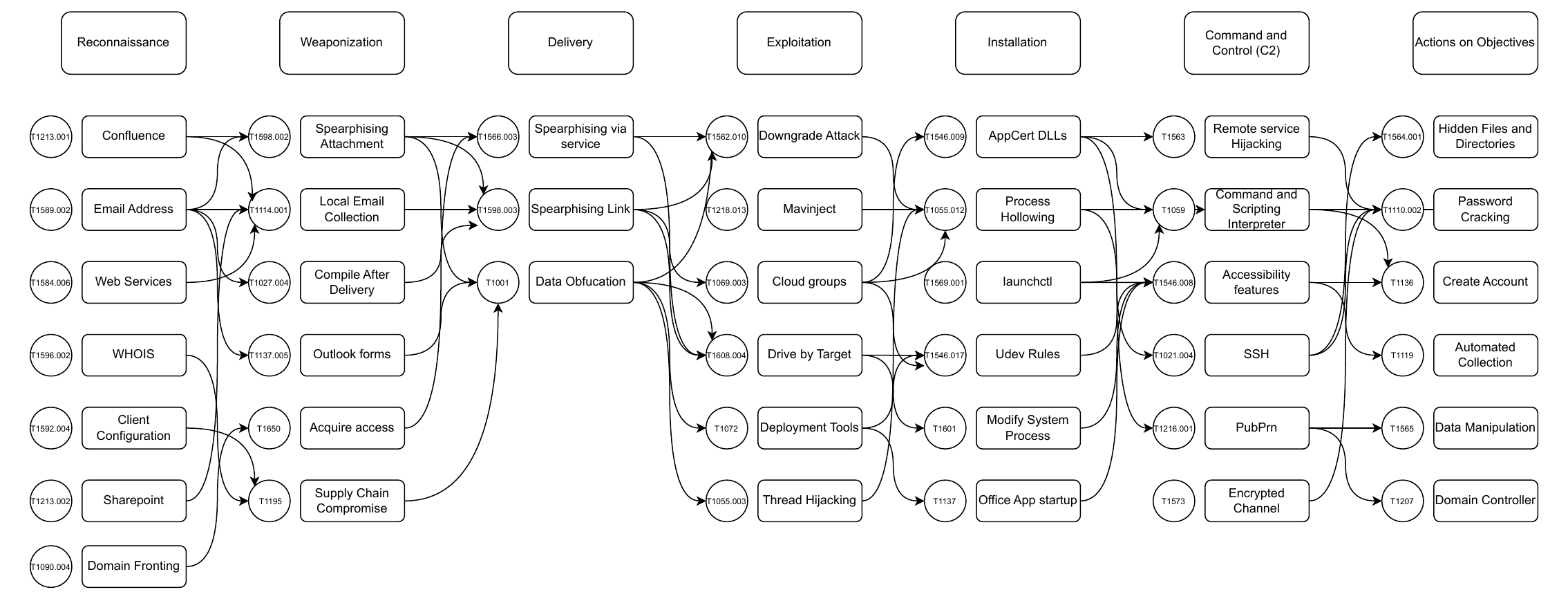}
    \caption{Kill Chain Graph generated from the narrative using
the trained model pipeline}
    \label{fig:killchain-graph}
\end{figure*}

\section{Results}

This section presents a comparative evaluation of the five classification models used to categorize MITRE ATT\&CK techniques into their respective Cyber Kill Chain (CKC) phases. Table~\ref{tab:ensemble_results} reports the performance of LightGBM (LGBM), Graph Neural Network (GNN), Transformer, BERT, and the proposed weighted soft voting ensemble across all seven CKC phases, using Accuracy, Precision, Recall, and F1-score as evaluation metrics. The results show that the weighted soft voting ensemble achieves the highest scores in most metrics across all phases, although the improvement over the GNN baseline is more subtle compared to previous iterations. Specifically, the F1-score advantage of the ensemble over GNN now lies between 0.03\% and 0.20\%, which, while numerically small, remains significant given the already high baseline performance where all F1-scores exceed 97\%. For example, in the \textit{Delivery} phase, the GNN achieves an F1-score of 99.28\%, with the ensemble improving this to 99.31\%, corresponding to a 0.03\% increase, while in the \textit{Exploitation} phase, the F1-score rises from 97.67\% to 97.87\% (a 0.2\% gain), and the largest observed improvement is in the \textit{Installation} phase, from 98.70\% to 98.83\% (a 0.13\% gain). These improvements occur consistently across all phases, confirming that the ensemble provides incremental gains even when the GNN already performs near optimally. Examination of the broader performance landscape shows that GNN remains the strongest individual model, consistently outperforming LGBM, Transformer, and BERT across all metrics and phases. LGBM generally ranks second, particularly in phases with clearer feature separability, while BERT performs better than Transformer due to its contextual semantic modeling, which is particularly beneficial in phases such as \textit{Exploitation} and \textit{Delivery}. The Transformer model remains the weakest across all phases, with accuracies ranging from 55.56\% in \textit{Delivery} to 86.81\% in \textit{Actions on Objectives}, suggesting that self-attention architectures without domain-specific adaptation may underperform in this task. The relatively narrow margin between the ensemble and GNN suggests that the latter already captures most discriminative patterns in the data; however, the ensemble still demonstrates measurable robustness by achieving higher or equal scores across all metrics and phases, benefiting from the architectural diversity of its constituent models. This diversity enables error correction in scenarios where a single model may fail, as predictions from weaker models can still contribute positively when combined with stronger models through weighted voting. From an operational cybersecurity perspective, even a 0.1\% to 0.2\% improvement in F1-score can represent a small but crucial set of correctly classified attack stages, leading to fewer false positives and false negatives and allowing Security Operations Center (SOC) analysts to allocate resources more effectively. In conclusion, the results in Table~\ref{tab:ensemble_results} confirm that while the GNN remains the most effective standalone classifier, the integration of multiple heterogeneous learners in a weighted soft voting framework produces a consistent performance uplift, reinforcing the value of ensemble learning in high-accuracy, high-stakes classification scenarios such as cyber threat detection.

To illustrate how our system performs in real-world conditions, we processed a detailed adversarial narrative describing a multi-stage cyber attack. The system takes this input text and maps it to the appropriate kill chain phases, generating a structured graph as output. The sample narrative used is as follows:

\begin{quote}
\small
"The adversary performed reconnaissance via subdomain enumeration and DNS zone transfers, uncovering a vulnerable webmail server. They delivered a phishing email to finance staff, containing a Word document weaponized with a VBA macro exploiting CVE-2017-0199. Upon opening, the macro executed PowerShell silently, installing a remote access trojan (RAT) that connected to a C2 server hosted on a compromised cloud instance. With access established, the attacker escalated privileges, moved laterally using stolen SMB credentials, and exfiltrated sensitive financial data over encrypted SFTP."

\end{quote}

This demonstration validates the model's ability to classify and sequence adversarial activities into coherent CKC phases. As visualized in Figure~\ref{fig:killchain-graph}, one possible interpretation of the attack chain begins with Reconnaissance through confluence-based intelligence gathering, followed by Weaponization involving the creation of a spear-phishing attachment. The Delivery phase includes sending a spear-phishing link, which transitions into Exploitation via a downgrade attack. Subsequently, Installation is achieved through process hollowing, enabling the adversary to maintain access. During the Command and Control stage, the attacker leverages a command and scripting interpreter to interact with the compromised system. Finally, the Actions on Objectives phase involves password cracking to achieve their end goals.

\section{Conclusion}
This study presented a novel machine learning framework for phase-wise prediction and semantic mapping of cyber kill chains by integrating MITRE ATT\&CK with Lockheed Martin’s cyber phase taxonomy. Semantic similarity via ATTACK-BERT and phase labeling yielded a comprehensive dataset covering seven kill chain stages. Leveraging ensemble learning with BERT variants, Transformers, and a GNN, the framework achieved strong predictive performance across all phases, with data augmentation mitigating class imbalance.

Although the ensemble model achieved superior accuracy, it increased computational complexity and inference time due to the need for predictions from multiple models. For critical tasks such as proactive threat forecasting, this trade-off is justified. The results validate the effectiveness of our multi-model approach, demonstrating that integrating diverse classifiers delivers tangible performance gains, culminating in a state-of-the-art predictive framework for Cyber Kill Chain mapping.Future work will focus on real-world validation through live threat intelligence integration and deployment within automated SOC pipelines.

\bibliographystyle{plain}
\bibliography{bibfile}

\IEEEpubidadjcol 

\end{document}